\documentclass[11pt]{article}
\usepackage{setspace}
\usepackage[latin9]{inputenc}
\usepackage{units}
\usepackage{amstext}
\usepackage{graphicx}
\usepackage{color}
\usepackage[round]{natbib}

\usepackage[T1]{fontenc}
\usepackage{ae,aecompl}

\usepackage{fancyhdr}
\bibpunct{(}{)}{;}{a}{}{,}

\makeatletter

\providecommand{\tabularnewline}{\\}

\@ifundefined{date}{}{\date{}}
\makeatother

\begin{document}

\title{Ligand-induced stabilization of the aptamer terminal helix in the
\textit{add} adenine riboswitch}

\author{Francesco Di Palma\,$^{1}$, Francesco Colizzi\,$^{1}$ and Giovanni
Bussi\,$^{1*}$}

\date{SISSA - Scuola Internazionale Superiore di Studi Avanzati, via Bonomea
265, 34136 Trieste, Italy}
\maketitle

\noindent Stabilization of the P1 stem in \textit{add} riboswitch \newline

\noindent keywords: P1 stem, RNA aptamer, adenine riboswitch, molecular dynamics simulation, free energy calculation.

{\renewcommand{\thefootnote}%
 {\fnsymbol{footnote}}
\footnotetext[1]{To whom correspondence should be addressed. Email: bussi@sissa.it;
Tel: +390403787407}}

\newpage
\noindent\textbf{\Large ABSTRACT}

\noindent Riboswitches are structured mRNA elements that modulate gene expression.
They undergo conformational changes triggered by highly-specific interactions
with sensed metabolites. Among the structural rearrangements engaged
by riboswitches, the forming and melting of the aptamer terminal helix,
the so-called P1 stem, is essential for genetic control. The structural
mechanisms by which this conformational change is modulated upon ligand
binding mostly remain to be elucidated. Here we used pulling molecular
dynamics simulations to study the thermodynamics of the P1 stem in
the \textit{add} adenine riboswitch. The P1 ligand-dependent stabilization
was quantified in terms of free energy and compared with thermodynamic
data. This comparison suggests a model for the aptamer folding in
which direct P1-ligand interactions play a minor role on the conformational switch when compared with those related to the ligand-induced aptamer preorganization.

\section{INTRODUCTION}

Riboswitches are ligand-responsive regulatory elements located in
untranslated regions of messenger RNAs \citep{serganov2013alexander}.
They change their conformation in response to specific metabolite
binding \citep{roth2009structural,edwards2010riboswitches,serganov2012metabolite}
and they have been proposed as modern descendants of an ancient sensory
and regulatory system in the RNA world \citep{breaker2012riboswitches}.
Many pathogenic bacteria use riboswitches to control essential metabolic
pathways and they are currently regarded as promising antibacterial
drug targets \citep{blount2006riboswitches,mulhbacher2010novel,deigan2011riboswitches}.
Riboswitches consist of an aptamer domain that binds the effector
ligand and an expression platform that transduces the ligand-induced
conformational ``switch'' into a modulation of gene expression \citep{barrick2007distributions,garst2009switch}.
Among more than 20 natural aptamer classes \citep{breaker2012riboswitches},
purine-sensing riboswitches have the peculiarity to recognize the
targeted purine by utilizing a conserved pyrimidine \citep{kim2008purine,batey2012structure}.
One of the most characterized members of this class is the adenine
sensing riboswitch (A-riboswitch) \emph{cis}-regulating the \emph{add} gene
in \emph{Vibrio vulnificus} \citep{mandal2003adenine}. The ligand-bound
structure of its aptamer is a three-way junction composed of three
stems (P1, P2, P3) with the ligand completely encapsulated into the
structure (Fig.~1) \citep{serganov2004structural,mandal2004gene}.
The specificity for adenine is ensured by canonical Watson-Crick (WC)
base pairing established between a uracil in conserved position and
the ligand \citep{noeske2005intermolecular,gilbert2006thermodynamic}.

The A-riboswitch acts as a translational regulator \citep{serganov2004structural,lemay2011comparative}.
In the absence of adenine the ribosome binding site and the initiation
codon, which are portions of the expression platform, are sequestered
by pairing with a portion of the aptamer (OFF-state, Fig.~2B).
The presence of adenine stabilizes an aptamer conformation in which
the terminal P1 helix is well structured and both the regulatory sequences
are available for ribosomal binding thus enabling mRNA translation
(ON-state, Fig.~2A) \citep{rieder2007ligand,lee2010real,leipply2011effects}.
The structural mechanism regulating the switch between the ON-
and the OFF-state upon ligand binding mostly remains to be elucidated.
The P1 stem is formed in the ON-state and it is disrupted in the OFF-state
\citep{mandal2003adenine,serganov2004structural}. 

It has been proposed that P1 is stabilized by the ligand \citep{batey2004structure} and
that this could be a common feature in many riboswitch classes \citep{montange2006structure}.
The role of ligand binding in the structural organization of the aptamer
has been investigated with single-molecule spectroscopy providing
an insightful overview on the folding dynamics \citep{neupane2011single},
yet lacking the critical atomistic details needed for an accurate structural characterization of the process {as extensively discussed by \cite{lin2012rna}.}
Although \emph{in silico} techniques
have been used to investigate the ligand role \citep{lin2008relative,sharma2009md,priyakumar2010role,gong2011role,allnr2013loop},
a quantitative estimation of the energetic contributions associated
to ligand binding, in particular regarding the role of direct P1-ligand
interactions, has not yet been provided.
In this context, state-of-the-art free-energy methods combined with atomistic simulations can bridge the gap providing an unparalleled perspective on the mechanism and
dynamics of the biomolecular process of interest \citep{dellago2009transition}.
In this work we used steered molecular dynamics (SMD) \citep{grubmuller1996ligand,sotomayor2007single}
simulations to study the thermodynamics of the P1 stem formation in
the presence and in the absence of the cognate ligand. We enforced the breaking of the P1 stem base pairs (bp) and then using a recently developed reweighting scheme \citep{colizzi2012rna} we quantitatively estimated the ligand-induced stabilization of the helix. The A9-U63 bp which directly stacks with adenine was used as a proxy for the P1 stability. Our non-equilibrium simulations provide measurements of the stability of the A9-U63 bp and quantify the direct
ligand-dependent stabilization of the pairing. In the following our results are presented and compared with melting and single-molecule experiments. A structural model for
the conformational switch emerging from the combination of our results
and previous experimental data is also discussed.

\section{RESULTS}

We carried out the simulations of the aptamer domain
of the \emph{add} A-riboswitch in different forms, namely the entire aptamer (PDB
id 1Y26) \citep{serganov2004structural} has been simulated in the presence (Holo) and in the absence (Apo) of the cognate ligand, the adenine; additionally, to better estimate the ligand-induced stabilization, we also simulated a truncated aptamer ($\triangle$1-8/64-71), both in the Apo and Holo forms. Long unbiased molecular dynamics (MD) for all the four systems were performed to test the stability of  the aptamer in different conditions. In the truncated systems the terminal bp was restrained in its initial configuration to mimic the presence of the rest of the stem. Furthermore, the full-length systems were pulled from the terminal bases to disrupt the entire P1 stem (Fig.~3) thus allowing its different stability between in the Holo and in the Apo forms to be qualitatively inferred. At last, to quantify this difference, SMD simulations of both the $\triangle$1-8/64-71 systems were done enforcing the breaking of the A9-U63 bp that directly stacks with the ligand (Fig.~4).

The stability of both the Apo and Holo systems was evaluated monitoring
the root mean square deviation (RMSD) from the native structure along 48 ns MD runs (Fig.~5A-B). Ligand removal (see Materials and Methods for details) did not affect the overall stability of the Apo aptamer in this time-scale, and secondary and tertiary structures were substantially unchanged.

The analysis of the trajectories obtained by pulling the P1 stem showed that the secondary and tertiary structure elements of the rest of the aptamer were not affected by the opening of the helix (data not shown). Focusing our attention on the P1 stem we observed that in the Apo form the A9-U63 bp (Fig.~3) was broken when the distance between the centers of mass of the terminal bases reached a value of $\approx$9.8 nm. Differently, in presence of the ligand a longer pulling was needed and the rupture only happened at a distance of $\approx$11.5 nm (Fig.~6). This behavior is compatible with the picture in which the ligand stabilizes the P1 stem \citep{montange2006structure}. It was however difficult to extract quantitative information on the ligand-P1 interaction from these simulations because the rupture is a stochastic event and extensive sampling would be required. 
{Moreover, as pointed out in a recent paper \citep{lin2012rna}, the end-to-end distance could be a non-optimal CV for pulling experiments or simulations
since local bp formation plays an
important role in global stem folding.}

The quantitative analysis of the P1-ligand interaction was better
obtained from the simulation of both the $\triangle$1-8/64-71 systems.
We verified that, when the P1 stem is replaced with the A9-U63 bp restrained to be in canonical WC pairing, the aptamer remains stable (Fig.~5C-D). Remarkably, fluorescence experiments have shown that the aptamer can also
fold and bind adenine when large fractions of the P1 stem are removed
\citep{lemay2007core}. This validates the possibility of using the two structures, $\triangle$1-8/64-71 Holo and $\triangle$1-8/64-71 Apo, to investigate the direct P1
stabilization given by the adenine. In the following we focus on the
SMD simulations performed on these truncated forms. Typical initial and final conformations from the SMD are shown in Fig.~4.

\subsection{Analysis of work profiles}

The unbinding event of the A9-U63 bp is described as a function
of the value of the steered RMSD in Fig.~7. The initial value corresponds to
the configuration with the WC pairing formed, whereas at the final
value (0.35 nm) the pairing is completely broken. Even if the ensembles
of work profiles for the two forms are broadly spread and superposed,
the free-energy profiles computed using the Jarzynski equality \citep{jarzynski1997nonequilibrium} as the exponential average of the
two sets of data are clearly distinguishable (Fig.~7A).
Qualitatively it is worth highlighting that during the breaking of
the A9-U63 bp the Apo form (red line) is always lower in free energy
than the Holo form (blue line). It follows that the breaking of the
monitored bp in the Apo form was unambiguously more probable than
in the Holo one (Fig.~7A). However, such an approach was still
a long way off from quantitatively accounting for the energetic stabilization
of the A9-U63 bp related to the presence of adenine in the binding
site. Within this framework there was no way to automatically detect
when the nucleobases reached the unbound configuration. It was thus difficult to avoid systematic errors in the comparison of the two systems. Furthermore, few low-work realizations occurred during the unpairing in the presence of adenine. In these low-work
realizations the number of hydrogen bonds was non-zero at large RMSD
values and the structural analysis of the trajectories revealed the
transient formation of a cis-sugar edge pair (data not shown) \citep{leontis2002non}.
Due to the exponential nature of the Jarzynski average, these low-work
realizations dominated the free energy profile for the Holo form further
compromising the possibility of a quantitative comparison with the
Apo form.

\subsection{Energetics of hydrogen bond breaking}

We thus analyzed the trajectories in terms of number of hydrogen bonds
formed between A9 and U63, a discrete variable that more strictly
reported on the breaking of the pairing. In this metrics, the bound
(1 or 2 hydrogen bonds) and unbound (0 hydrogen bond) ensembles could
be clearly and unambiguously identified thus allowing a quantitative
comparison between the Apo and the Holo system. Additionally, the
configurations from the outlier trajectories could be assigned properly
to one or the other ensemble in spite of their atypical RMSD value.

The differences in free energy ($\Delta F$) between the ensembles,
with and without hydrogen bonds, was computed using a reweighting scheme \citep{colizzi2012rna}. The values and the associated
standard errors were estimated for both systems. For the Apo form
$\Delta F=-2.5\pm1.4\ kJ/mol$ suggesting that the bp could spontaneously
break in the absence of adenine. For the Holo form $\Delta F=1.9\pm1.7\ kJ/mol$
implying that the presence of the ligand and its pairing with U63
stabilized the stacked bp. The $\Delta\Delta F$ between the two forms
is equal to $-4.4\pm2\ kJ/mol$. This value quantifies the thermodynamic
stabilization to the formation of the base pair which directly interacts
with adenine in the P1 stem.

\section{DISCUSSION AND CONCLUSIONS}

Our simulations at atomistic detail provide for the first
time the free-energy contribution of ligand stacking to the formation
of the P1 stem in a riboswitch. In particular, the presented \emph{in
silico} approach allows the energetics involved in the aptamer stabilization
upon ligand binding to be dissected in detail. Below we compare our
results with single-molecule manipulation, both \emph{in vitro} and
\emph{in silico}, and thermodynamic data from dsRNA melting experiments.
We also provide a model for ligand-modulated co-transcriptional folding
of the \emph{add} riboswitch.

\subsection{Comparison with related works}

Our results are in nice agreement with thermodynamic data based on
dsRNA melting experiments \citep{mathews2004incorporating,turner2010nndb}.
The comparison between our simulations and those experiments can be
straightforwardly achieved by considering the pairing between U62
and the sensed adenine as an additional terminal bp of the P1 stem.
The direct stabilization of the P1 stem due to the cognate-ligand
binding should be thus equivalent to that given by adding one further
AU bp to the P1 helix. Using the most recent nearest neighbor energy
parameters for the comparison of RNA secondary structures \citep{mathews2004incorporating,turner2010nndb},
the free-energy difference between the sequence of the P1 stem with
and without the additional AU base pair, %
\begin{tabular}{c}
$_{\textrm{5'-CGCUUCAUA\emph{A}-3'}}$\tabularnewline
$^{\textrm{3'-GUGAAGUAU\emph{U}-5'}}$\tabularnewline
\end{tabular}and %
\begin{tabular}{c}
$_{\textrm{5'-CGCUUCAUA-3'}}$\tabularnewline
$^{\textrm{3'-GUGAAGUAU-5'}}$\tabularnewline
\end{tabular}, can be computed \citep{hofacker1994fast,lorenz2011viennarna} as
$\Delta\Delta F=-3.7\ kJ/mol$, consistently with our results.

Our free-energy estimates complement previously reported investigations
in which the role of the ligand in the folding process of the A-riboswitch
has always been referred to the whole aptamer \citep{lin2008relative,neupane2011single}
and never specifically to the P1 stem. Using a one-bead-per-nucleotide
coarse-grained model, the $\Delta\Delta F$ has been computed as approximately
$-15\ kJ/mol$ \citep{lin2008relative}.
{Notably also this calculation has been done using a shortened P1 stem,
possibly affecting the $\Delta F$ estimation.}
Single-molecule force spectroscopy
experiments have been also performed to characterize the folding pathway
of the aptamer with an estimated $\Delta\Delta F\simeq-33\ kJ/mol$
\citep{neupane2011single}. However in both these works the separated contributions
of the P1-ligand stacking, of the interaction between the ligand and
the junctions J1-2, J2-3 and J3-1, and of the interaction between
loops L2-L3 could not be discerned (secondary structure elements labeled
as in Fig.~8).

From the comparison of our data with the above mentioned experimental
and computational works, a twofold modular role for the ligand emerges.
On the one hand, the binding of adenine can contribute to the aptamer
preorganization and it could allow the long-range induction of the
tight hydrogen-bonding and base-stacking networks observed in the
native state \citep{rieder2007ligand,lee2010real}. This preorganization
would reduce the distance between A9 and U63, thus increasing the
probability of their pairing. A similar mechanism has been proposed
also for the SAM-I riboswitch \citep{whitford2009nonlocal}. On the
other hand, adenine binding enhances the P1 formation by direct stacking
interaction, mimicking the extension of the helix by an additional
bp. Notably, the energetic contribution of the direct stacking is
smaller than that involved in the aptamer preorganization. The latter can be estimated as the difference between the global ligand-induced aptamer stabilization \citep{lin2008relative,neupane2011single} and the stacking contribution dissected in our work.

\subsection{Folding model}

Our work provides atomistic details and energetic estimates to the
currently accepted model for the folding of the \emph{add} riboswitch
upon ligand binding \citep{rieder2007ligand,lee2010real,leipply2011effects}.
Altogether, our data and the related experimental works suggest a
folding model as depicted in Fig.~8. Initially, only the
P2 and P3 stems and the corresponding loops (L2, L3, still not interacting
each other) are formed and not fully stable (Fig.~8A). Then, adenine binding
allows for a preorganization of the aptamer where the three junctions
arrange around the ligand (Fig.~8$\textrm{B}_{\textrm{1}}$), stabilizing also the previously formed
helices \citep{rieder2007ligand}. It has not been established clear if the loop-loop
interaction is formed before or after ligand binding \citep{leipply2011effects}.
Thus, an alternative pathway, the junctions and the P1 could acquire
a partially folded conformation also in the absence of adenine (Fig.~8$\textrm{B}_{\textrm{2}}$) \citep{lee2010real}.
Finally, the P1 helix becomes fully structured and stabilized by the
ligand (Fig.~8C), to the detriment of the expression platform (see Fig.~2)
\citep{lemay2011comparative}. This step is mandatory for translation
to be initiated. We quantified the ligand contribution to the P1 stem
formation due to direct interactions to be approximately -4 kJ/mol.

Our result is compatible with both the folding pathways (Figs.~8$\textrm{B}_{\textrm{1}}$
and 8$\textrm{B}_{\textrm{2}}$) irrespectively of their relative
population and cannot discriminate among them. The relative probability
between the two paths can be modulated by the ligand concentration
and its binding affinity. On the one hand, the intermediate shown
in Fig.~8$\textrm{B}_{\textrm{1}}$ could be relevant for ligand-RNA
binding in an early transcriptional context in which the last 9 nucleobases
(i.e those allowing P1 formation) of the aptamer have not yet been
synthesized. Indeed, it has been shown that also an aptamer missing
a large portion of the P1 stem is able to bind adenine \citep{lemay2007core}.
On the other hand, the intermediate shown in Fig.~8$\textrm{B}_{\textrm{2}}$
could be populated at low-ligand concentration once the nucleobases
allowing P1 formation are synthesized. Later on, after the synthesis
of the expression platform, ligand binding could shift the thermodynamic
equilibrium towards one of two competing riboswitch conformations
(P1 formed and non formed).

\subsection{Conclusion}

Ligand-induced stabilization of the P1 stem is crucial for A-riboswitch
regulation and function. Here we quantified the direct interaction
between adenine and P1 stem and analyzed it at atomistic detail. Our
results suggest a model for the aptamer folding in which the direct
P1-ligand interactions play a minor role on the conformational switch when compared with those
related to the ligand-induced aptamer preorganization. Because the
structural/functional role of the aptamer terminal helix is a common
feature in the ``straight junctional'' riboswitches \citep{serganov2013alexander},
we foresee a wider validity of the model presented herein.

\section{MATERIALS AND METHODS}

\subsection{System description and set-up}

We simulated the Holo and the Apo form of the A-riboswitch aptamer
domain both composed of 71 nucleotides. The Apo form was generated
by adenine removal from the ligand-bound (Holo) crystal structure
(PDB id: 1Y26) \citep{serganov2004structural}. This deletion is justified
by the fact that the Apo and Holo form have been shown experimentally
to share an overall similar secondary structure \citep{lemay2011comparative}.
This is at variance with e.g. the \emph{pbuE} adenine riboswitch
in which the two structures are different. The generation of
the Apo form by simply removing the ligand has been adopted also in
a recent work \citep{allnr2013loop}. Molecular dynamics (MD) simulations
were performed using the Amber99 force field \citep{wang2000well}
refined with the \emph{parmbsc0} corrections \citep{perez2007refinement}. From the analysis of the SMD trajectories we do not expect the refinement on the $\chi$ torsional parameters \citep{zgarbova2011refinement} to affect the results.
Adenine was parametrized using the general Amber force field (gaff)
\citep{wang2004development}. Partial atomic charges were assigned
using the restricted electrostatic potential fit method \citep{bayly1993well}
based on an electronic structure calculation at the HF/6-31G{*} level
of theory performed with Gaussian03 \citep{g03}. Bond-lengths were constrained with
LINCS \citep{hess1997lincs} and the electrostatic interactions were
calculated using the particle-mesh Ewald method \citep{darden1993particle}.
For both forms, the following protocol was used to prepare the systems
(Table \ref{tabS1} for details) for MD simulations: steepest descent minimization (200 steps) starting from the X-ray structure. Solvation with $\approx$13000 TIP3P water molecules \citep{jorgensen1983comparison}
and NaCl at 0.15 M concentration (plus extra Na$^{+}$ counter-ions to neutralize the
charges of the systems) in a hexagonal prism
(lattice vectors in nm {[}(10,0,0), (0,7,0), (0,$\frac{7}{2}$,$\frac{7\cdot\sqrt{3}}{2}$){]})
that was created orienting the major length of the aptamer along the X axis. Steepest descent minimization (200 steps) for ions and solvent; the systems were thermalized at 300 K, initially for 200 ps with frozen
solute positions and then for 5 ns in NPT ensemble (1 atm) with stochastic velocity rescaling \citep{bussi2007canonical} and Berendsen barostat \citep{berendsen1984molecular}; to maintain the systems oriented along the largest lattice vector (X) a restraint was imposed
with a force constant of 4$\cdot$10\textsuperscript{3}$\unitfrac{(\unitfrac{kJ}{mol})}{nm^{2}}$
on the Y and Z components of the distance between phosphate atoms of A52 and G71. Each system was simulated for 48 ns in NVT ensemble to assess the stability of the aptamer.

\subsection{Steered molecular dynamics}

To perform SMD simulations inducing the opening of the whole P1 stem the systems were solvated again with $\approx$39500 water molecules in a larger rhombic dodecahedral box with distance between periodic images equal to 12 nm, adding ions to maintain the same ionic strengh (P1-SMD systems in Table \ref{tabS1}). The same protocol described above was applied for the minimization, thermalization and equilibration of this larger Holo and Apo systems for the pulling simulations. An incremental separation between the centers
of mass of the terminal nucleotides (C1 and G71) was imposed from
an initial value of 1.05 nm to a value sufficient to completely unfold the 9 bp of
the P1 helix (Apo 10.05 nm, Holo 11.75 nm) at a speed of 0.56 nm/ns
(see Fig.~3). The spring constant was set to 3.9$\cdot$10\textsuperscript{4}$\unitfrac{(\unitfrac{kJ}{mol})}{nm^{2}}$.

The first eight bp of the P1 stem (i.e. whole stem from C1 to U8 and from A64 to G71,  except for A9-U63 bp) were then cut in both systems creating the $\triangle$1-8/64-71
Holo and $\triangle$1-8/64-71 Apo structures (Table \ref{tabS1} for details). Water molecules were allowed to relax filling the space left
by the 16 removed bases through an additional 1 ns equilibration in which
the positions of aptamer atoms were frozen followed by 5 ns of unrestrained NPT simulation. Then the systems were simulated for 48 ns in the NVT ensemble restraining the terminal bases in the initial state to avoid any spontaneous flipping. The pairs deletion is not biologically meaningless, because it has been shown experimentally that a series of aptamer variants with shorter P1 helix are still able to bind the ligand \citep{lemay2007core}. The deletion reduced the noise during the pulling allowing to focus the calculation on the influence of the ligand on the A9-U63 pairing. This bp rupture was here enforced by pulling on the RMSD between the heavy atoms of A9 and U63 with reference to the crystal structure. This collective variable (CV) was chosen as it identifies the native conformation (RMSD $\approx$ 0) of the A9-U63
bp, which is necessary for the initiation of the P1 stem formation. The steered CV was pulled at constant velocity of 0.175 \unitfrac{nm}{ns} from 0 to 0.35 nm in 2 ns. This pulling induced the complete opening of the A9-U63 bp in presence and absence of the ligand (Fig.~4). The spring constant was set to 3.9$\cdot$10\textsuperscript{4}$\unitfrac{(\unitfrac{kJ}{mol})}{nm^{2}}$. The starting points were extracted equidistantly (one every 16 ps) from a 8.192 ns run NVT ensemble restraining the RMSD value of those atoms at 0. For the two system 512 independent SMD simulations were performed, corresponding to an aggregate time of approximately 1$\mu$s each. Simulations were carried out with the Gromacs 4.0.7 program package \citep{hess2008gromacs} combined with the PLUMED 1.3 plug-in \citep{bonomi2009plumed}.

\subsection{Analysis}

The Jarzynski equality \citep{jarzynski1997nonequilibrium} was used
to estimate the equilibrium free-energy landscape of the process from
the collected work profiles. The simulations were then analyzed using
a recently proposed scheme \citep{colizzi2012rna} which combines an
identity by Jarzynski \citep{PhysRevE.56.5018} with the weighted-histogram
analysis method \citep{kumar1992weighted}. The algorithm allows the
free-energy profiles to be projected onto any \emph{a posteriori}
chosen CV. {It is well known that free-energy calculations using Jarzynski-based relationships are difficult to converge. Statistical errors were thus estimated by the bootstrapping procedure described in \cite{do2013rna} indicating that our results were converged within $\approx k_BT$.} The VIENNA RNA
package \citep{hofacker1994fast,lorenz2011viennarna} was used to compare
our results with the thermodynamic data based on dsRNA melting experiments
\citep{mathews2004incorporating,turner2010nndb}.

\section{ACKNOWLEDGMENTS}

{

We thank Daniel Lafontaine and Gabriele Varani for reading the manuscript and providing critical comments.
Sandro Bottaro is also
acknowledged for carefully reading the manuscript
and suggesting several improvements.
We acknowledge the CINECA award no.
HP10B2G6OF, 2012 under the ISCRA initiative for the availability of high performance computing resources.
}
The research leading to these results has received funding from the European Research Council under the European Union's Seventh Framework
Programme (FP/2007-2013) / ERC Grant Agreement n. 306662, S-RNA-S.

\newpage
\begin{table}[!h] \caption{Specifications for the simulated systems} \begin{tabular}{ccccc} \hline  System & Total atoms & Water molecules & aptamer atoms & Ions (Na+Cl)\tabularnewline \hline  Holo form & 41628  & 13078  & 2257 & 122 (96+26)\tabularnewline \hline  Apo form & 41676 & 13099 & 2257 & 122 (96+26)\tabularnewline \hline  Holo (P1-SMD) & 120654 & 39364 & 2257 & 290 (180+110)\tabularnewline \hline  Apo (P1-SMD) &120588 & 39347 & 2257 & 290 (180+110)\tabularnewline \hline Holo $\Delta$1-8/64-71 & 41101 & 13078 & 1746 & 106 (80+26)\tabularnewline \hline  Apo $\Delta$1-8/64-71 & 41149 & 13099 & 1746 & 106 (80+26)\tabularnewline \hline \end{tabular}
\label{tabS1} \end{table}

\newpage

\begin{figure}[!h]
\begin{centering}
\includegraphics[width=8cm]{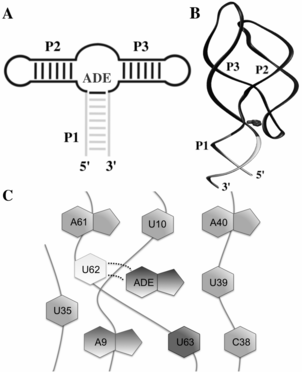} 
\par\end{centering}
\caption{
Adenine riboswitch aptamer and binding site. A) Secondary structure elements and B) 3-dimensional structure with bound adenine. The P1 stem is grey, the other stems and loops are black. C) Cartoon representation of the binding site; the two dotted lines represent the hydrogen bonds of the WC pairing between the U62 and the ligand.
}
\label{fig1} 
\end{figure}

\newpage
\begin{figure}[!h]
\begin{centering}
\includegraphics[width=8cm]{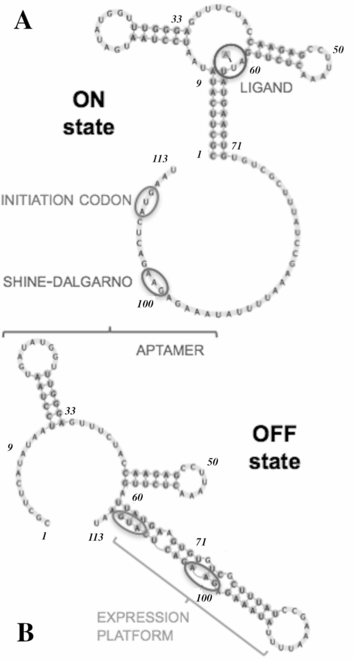} 
\par\end{centering}
\caption{
Secondary structure representation of the \emph{add} riboswitch in the ON (A) and OFF (B) states. The ligand, the initiation codon and the Shine-Dalgarno sequence are labeled.}
\label{fig3new} 
\end{figure}

\newpage
\begin{figure}[!h]
\begin{centering}
\includegraphics[width=8cm]{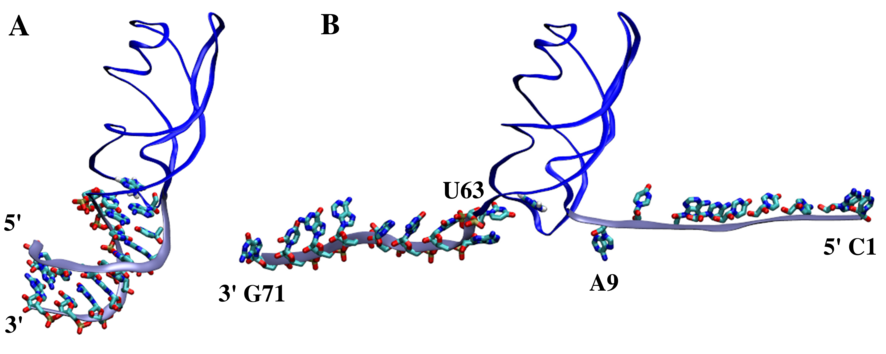}
\par\end{centering}
\caption{
Initial (A) and final (B) configuration of the SMD
simulation opening the P1 stem here shown for the Holo form. The backbone of the
aptamer is in blue except for the P1 stem, in light blue. The ligand
and the 18 bases forming the helix are shown. The P1 stem is formed
in A and disrupted in B.
}
\end{figure}

\newpage
\begin{figure}[!h]
\begin{centering}
\includegraphics[width=8cm]{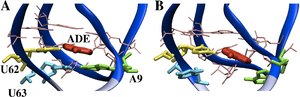}
\par\end{centering}
\caption{
Representative structures of the Holo binding pocket at the beginning
(A, RMSD = 0 nm) and at the end (B, RMSD = 0.35 nm) of the SMD. The portion of the P1 stem removed in our simulations is in light blue. Bases forming
the binding pocket are labeled, ligand is shown in red. A9-U63 pair
is formed in A and disrupted in B.
}
\end{figure}

\newpage
\begin{figure}[!h]
\begin{centering}
\includegraphics[width=8cm]{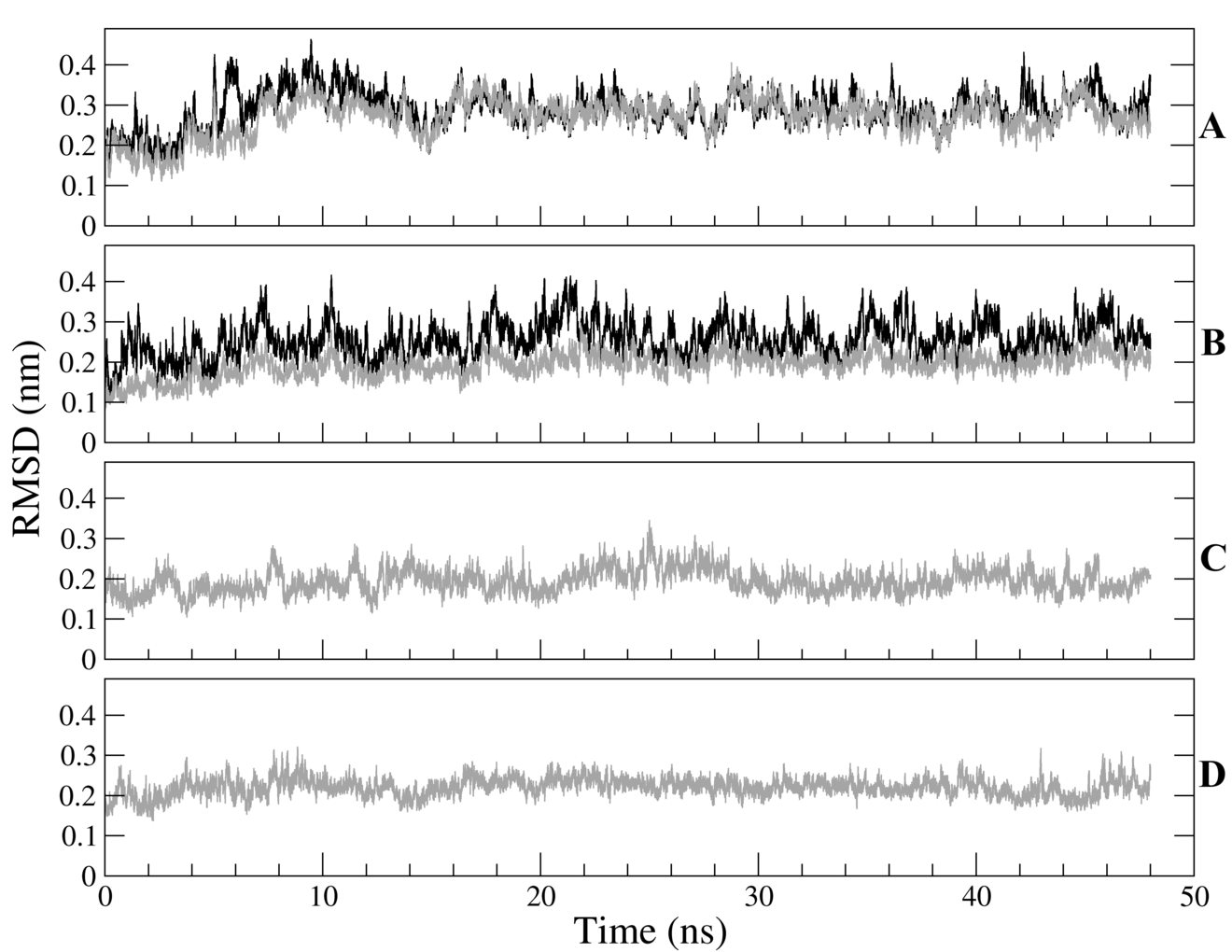}
\par\end{centering}
\caption{
Root mean square deviation (RMSD) from native structure.
A) Holo form during 48 ns equilibration, computed
on the whole aptamer (black) and on the bases from 9 to 63 (grey).
B) Same as A) done on the Apo form (whole aptamer,
black; bases from 9 to 63, grey). The difference between black and grey profiles in both panels, indicates that the P1 stem is less stable than the rest of the aptamer.
C) $\triangle$1-8/64-71 Holo RMSD along the NVT 48 ns equilibration. D) Same as C) for the $\triangle$1-8/64-71 Apo form.
}
\end{figure}

\newpage
\begin{figure}[!h]
\begin{centering}
\includegraphics[width=8cm]{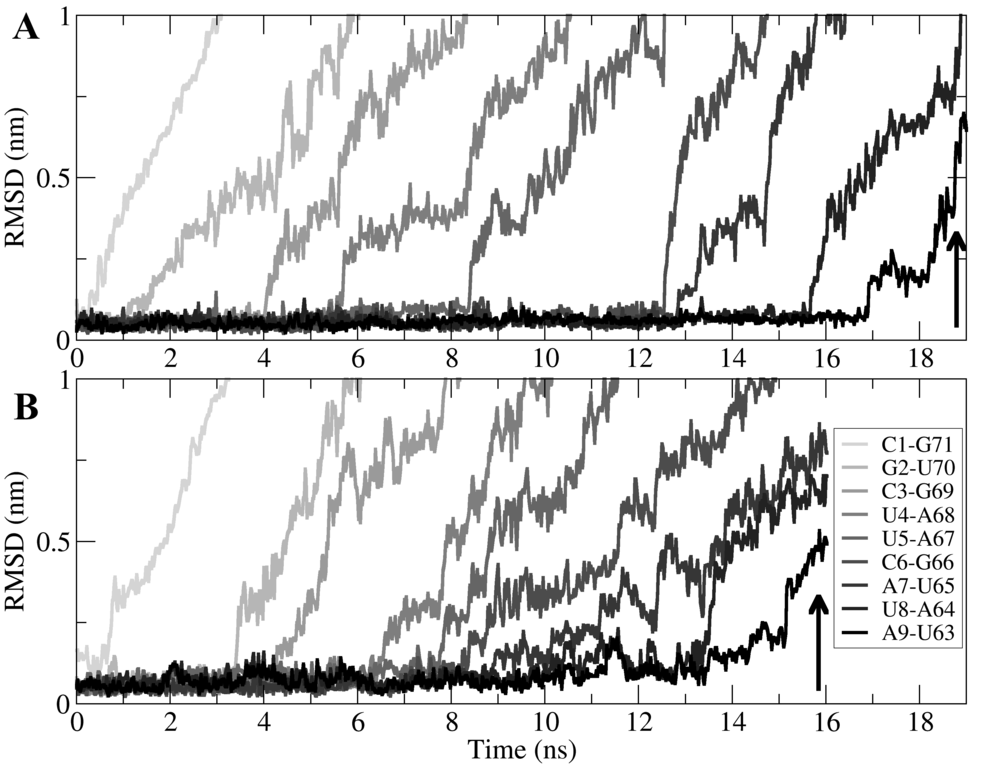}
\par\end{centering}
\caption{
Base-pair ruptures during P1 pulling. In the pulling
simulations the 9 bp forming the P1 stem were unpaired. We here monitored
the RMSD of each bp (grey-scale) from their native conformation (A, Holo; B, Apo). A9-U63 bp (in black) was disrupted (RMSD $\approx$0.5 nm, arrows) later in the Holo form ($\approx$19 ns) than in the Apo one ($\approx$16 ns).
}
\end{figure}

\newpage
\begin{figure}[!h]
\begin{centering}
\includegraphics[width=8cm]{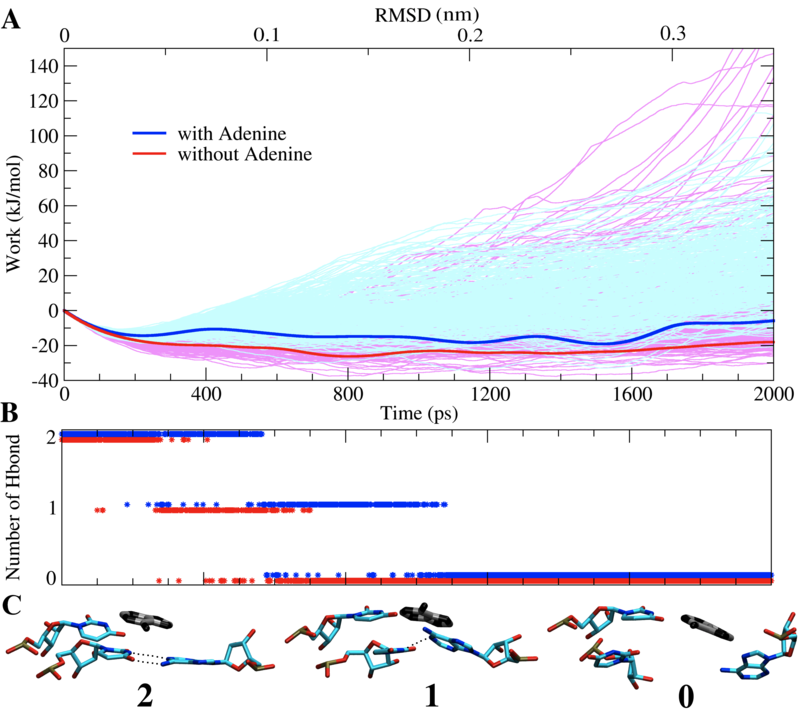}
\par\end{centering}
\caption{
Unbinding process of the A9-U63 bp. (A) Mechanical work performed
as a function of the value of the steered RMSD, or equivalently of
time, for 512 simulations for Apo (pink) and Holo form (light-blue).
The respective free energies resulting from the Jarzynski equality
are shown in thicker red and blue lines. The initial free-energy decrease
related to the entropy gain induced by the restraint movement has
no consequence on the final result. (B) Hydrogen bonds occurrence
for both the systems Apo (red) and Holo (blue). (C) Snapshots of the
Holo form (ligand in black) with 2, 1 or 0 hydrogen bond (dotted lines)
formed between A9-U63.
}
\end{figure}

\newpage
\begin{figure}[!h]
\begin{centering}
\includegraphics[width=8cm]{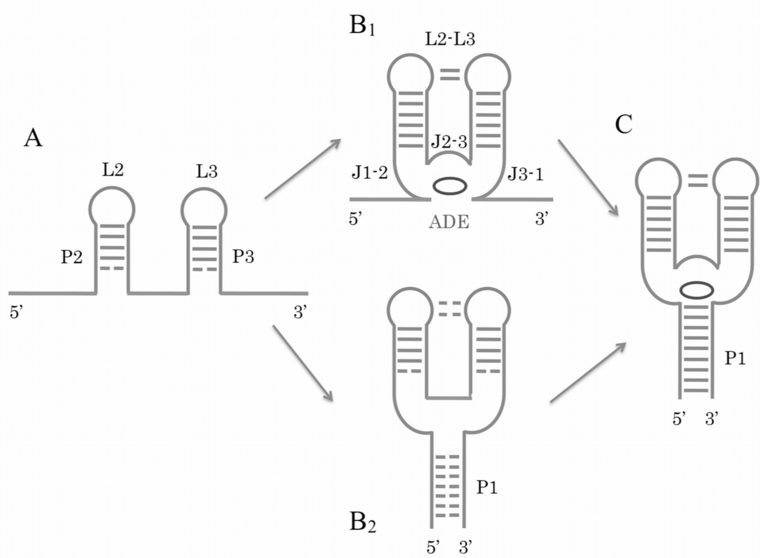}
\par\end{centering}
\caption{
Schematic representation of the aptamer secondary structure in its
folding intermediates with and without the ligand. A) Stems P2 and
P3, loops L2 and L3 are folded but not stable. $\textrm{B}_{\textrm{1}})$
The junctions (J1-2, J2-3, J3-1) arrange around the ligand (ADE) and
the inter-loop pairings occur (L2-L3) stabilizing also the stems.
$\textrm{B}_{\textrm{2}})$ Alternative possible intermediate in which all
the 3 stems are not completely and stably folded before ligand binding.
The junctions and the tertiary interaction between the loops are not
stable. C) P1 stem is fully folded and stabilized by the ligand.
}
\end{figure}

\end{document}